\documentclass[10pt, a4paper]{article}

\usepackage{lrec-coling2024} 

\usepackage{listings}

\usepackage{natbib}
\usepackage{multibib}
\makeatletter
\def\@mb@citenamelist{cite,citep,citet,citealp,citealt,citepalias,citetalias}
\makeatother
\newcites{languageresource}{~}

\usepackage{graphicx}
\usepackage{tabularx}
\usepackage{soul}

\usepackage{xcolor}
\usepackage{hyperref}
 \definecolor{darkblue}{rgb}{0, 0, 0.5}
  \hypersetup{colorlinks=true, citecolor=darkblue, linkcolor=darkblue, urlcolor=darkblue}

\usepackage{xstring}

\usepackage{color}

\usepackage{soul}
\usepackage{algorithm2e}
\usepackage{tabularx}
\usepackage{booktabs}
\usepackage{multirow}
\usepackage{amsmath}
\usepackage{float}
\usepackage{algpseudocode}
\usepackage{caption}
\usepackage{afterpage}
\usepackage{float}
\usepackage{hyperref}
\usepackage{listings}

\definecolor{light-gray}{gray}{0.95}
\definecolor{olive-green}{RGB}{34, 139, 34}
\definecolor{dark-orange}{RGB}{255, 140, 100}

\lstdefinestyle{pythonstyle}{
    backgroundcolor=\color{light-gray},
    commentstyle=\color{olive-green},
    keywordstyle=\color{black},
    numberstyle=\tiny\color{black},
    stringstyle=\color{dark-orange},
    basicstyle=\ttfamily\small,
    breakatwhitespace=false,
    breaklines=true,
    captionpos=b,
    keepspaces=true,
    numbers=left,
    numbersep=5pt,
    showspaces=false,
    showstringspaces=false,
    showtabs=false,
    tabsize=2,
}

\title{LocalTweets to LocalHealth: A Mental Health Surveillance Framework Based on Twitter Data}

\name{
    \begin{tabular}[t]{c}
        Vijeta Deshpande$^{1}$, 
        Minhwa Lee$^{2}$, 
        Zonghai Yao$^{3}$, 
        Zihao Zhang$^{3}$, \\
        Jason Brian Gibbons$^{4}$, 
        Hong Yu$^{1, 3, 5}$
    \end{tabular}
} 

\address{
    ${^1}$University of Massachusetts Lowell, \\
    ${^2}$University of Minnesota, \\
    ${^3}$University of Massachusetts Amherst, \\
    ${^4}$University of Colorado Anschutz Medical Campus, \\
    ${^5}$University of Massachusetts, Chan Medical School,\\
    vijeta\_deshpande@student.uml.edu, lee03533@umn.edu, \{zonghaiyao, zihaozhang\}@umass.edu, \\
    jason.gibbons@cuanschutz.edu, hong\_yu@uml.edu
}

\abstract{
Prior research on Twitter (now X) data has provided positive evidence of its utility in developing supplementary health surveillance systems. In this study, we present a new framework to surveil public health, focusing on mental health (MH) outcomes. We hypothesize that locally posted tweets are indicative of local MH outcomes and collect tweets posted from 765 neighborhoods (census block groups) in the USA. We pair these tweets from each neighborhood with the corresponding MH outcome reported by the Center for Disease Control (CDC) to create a benchmark dataset, LocalTweets. With LocalTweets, we present the first population-level evaluation task for Twitter-based MH surveillance systems.
We then develop an efficient and effective method, LocalHealth, for predicting MH outcomes based on LocalTweets. When used with GPT3.5, LocalHealth achieves the highest F1-score and accuracy of 0.7429 and 79.78\%, respectively, a 59\% improvement in F1-score over the GPT3.5 in zero-shot setting. We also utilize LocalHealth to extrapolate CDC's estimates to proxy unreported neighborhoods, achieving an F1-score of 0.7291. Our work suggests that Twitter data can be effectively leveraged to simulate neighborhood-level MH outcomes.\thanks{Following responsible data practices, we will share data for requests that align with our privacy policy. Corresponding Author: \href{vijeta_deshpande@student.uml.edu}{vijeta\_deshpande@student.uml.edu}} \newline  }

\begin{document}

\maketitleabstract

\section{Introduction}
For effective design of public health interventions, it is critical to have surveillance systems that are reliable and fast-acting. 
Traditional health surveillance systems often resort to survey-based reporting of health outcomes hence, are subject to response bias, and significant temporal lag \cite{bitsko2022mental}. For the timely design and implementation of health intervention programs, real-time data monitoring, processing, and estimation systems are required \cite{simonsen2016infectious}. Electronic Health Records (EHR) based surveillance systems carry the potential to overcome the disadvantages of the traditional systems \cite{greco2023transformer, simonsen2016infectious}. While EHRs offer valuable insights, operational challenges, their expensive nature, and relatively delayed updates in information compared to social media platforms reduce their effectiveness for real-time public health surveillance \cite{kataria2020electronic,menachemi2011benefits}. Hence, the exploration of supplementary data sources for health surveillance is needed.

\begin{figure*}[t] 
    \centering
    \includegraphics[width=0.8\textwidth]{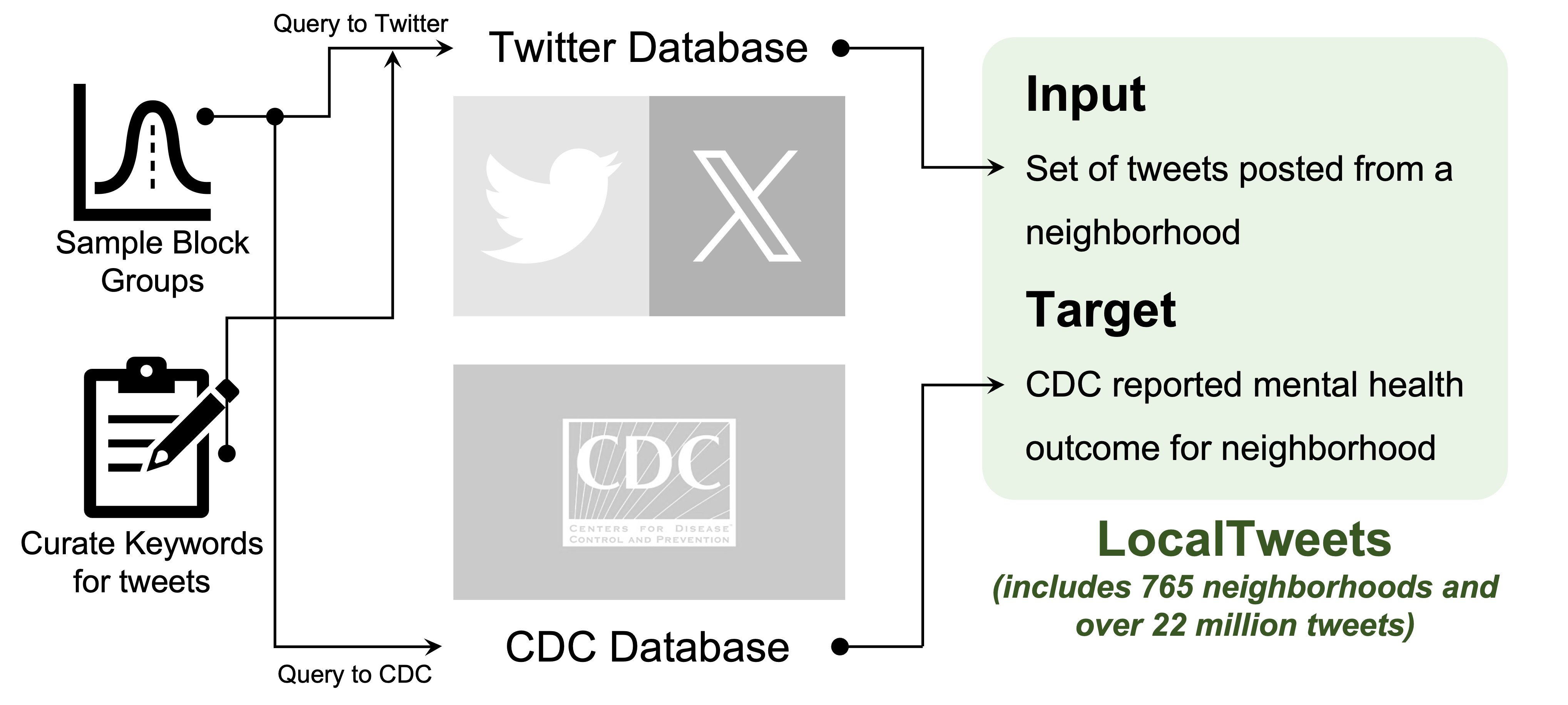}
    \caption{\textbf{Data Collection Process.} In this figure, we present a simple schematic of our data curation process. First, we sample 1K neighborhoods (i.e., block groups or BGs) and curate a list of keywords for three categories of tweets, to form queries. Secondly, we query the CDC and Twitter databases to collect the desired data. Lastly, for each BG, we join the set of tweets posted from the BG with the reported health outcome from the CDC database. The final cleaned version of LocalTweets includes 765 unique BGs, spans over five years, and includes over 22 million tweets.}
    \label{fig:data_collection_flowchart}
\end{figure*}

Social media platforms as a data source are proving to be important for various surveillance applications \cite{shakeri2021digital}, with Twitter (now X\footnote{We refer to X by its older name `Twitter' and refer to the messages posted on X as tweets.}) being one of the most explored platforms for population health surveillance applications \cite{greco2023transformer, mavragani2020infodemiology, jordan2018using, pilipiec2023surveillance}. The previous decade evidenced a spectrum of research efforts to highlight the utility of Twitter data for health surveillance \cite{coppersmith2015adhd, naseem2022benchmarking, nguyen2017geotagged, athanasiou2023long, klein2022toward, coppersmith2014quantifying, shakeri2021digital}. Numerous studies conducted correlation analysis to emphasize that Twitter activities are highly correlated with the reported outcomes, at the national, state, and even county level \cite{coppersmith2015adhd, coppersmith2014quantifying, paul2011you}. Several studies developed Twitter surveillance systems with advanced Natural Language Processing (NLP) methods to identify tweets indicating serious health concerns \cite{naseem2022benchmarking, barbieri2020tweeteval, rosenthal2019semeval, yadav2020identifying}. However, the research efforts to develop population-level health outcome prediction systems have been quite limited \cite{nguyen2017geotagged, nguyen2016building, nguyen2017social, nguyen2016leveraging, wang2020regional, athanasiou2023long}. Recently conducted studies by \citet{barbieri2020tweeteval, naseem2022benchmarking} show that the population-level inference tasks are absent in the current collection of Twitter evaluation benchmarks. 

Exacerbating the issue, several analytical limitations within the population-level studies constrain the transferability of findings. First, keyword-based tweet filtering is hypothesized to improve prediction systems, but this has not been tested. Second, studies have focused on larger geographical areas (census tract being the smallest area considered in \citet{nguyen2016building}) hence, indirectly normalizing worsened health conditions in smaller, resource-deprived areas. Lastly, most population-level prediction systems presented in literature employ rule-based or count-based feature extraction to encode tweets, which lacks the benefits of advanced pre-trained language models.

To overcome the above-mentioned limitations in the literature, we first present a benchmark dataset; LocalTweets; that enables the \textbf{prediction of neighborhood-level mental health (MH) outcomes}\footnote{For a precise definition of the MH outcome refer to \cite{places2023def, centers2022places}} from locally posted tweets. Compared to previous studies we \textbf{focus on a much smaller geographical unit, Census Block Group (BG)} \footnote{United States Census Bureau has defined geographical units to collect data from. Census Block Groups are areas with a population ranging from 600 to 3000. More detailed definitions can be found at \citet{bg2023def}} in LocalTweets, refer to Figure \ref{fig:data_collection_flowchart}. LocalTweets includes data for 765 unique BGs, spans over a period of five years (2015-2019), and includes more than 22 million tweets. Furthermore, we propose an efficient and effective analytical framework, LocalHealth, that \textbf{leverages language models to encode locally posted tweets} and predicts MH outcomes based on the encodings. We evaluate LocalHealth with extensive experiments and find that the \textbf{unfiltered tweets present better generalization properties} compared to filtered tweets (containing MH-related keywords). With LocalHealth we achieve an F1-score of 0.7429 in predicting future outcomes and 0.7291 in predicting outcomes for a proxy set of unreported BGs. 

Our work thus lays the groundwork for the development of a neighborhood-level, real-time MH surveillance system and holds substantial benefits for public health decision-making. 
For example, the presented work in this study can directly be used to identify neighborhoods that can benefit from additional MH care resources and the establishment of community MH programs. In the following sections, we delineate details of our analysis.

\section{Related Works}

\begin{figure*}[ht!] 
    \centering
    \includegraphics[width=0.9\textwidth]{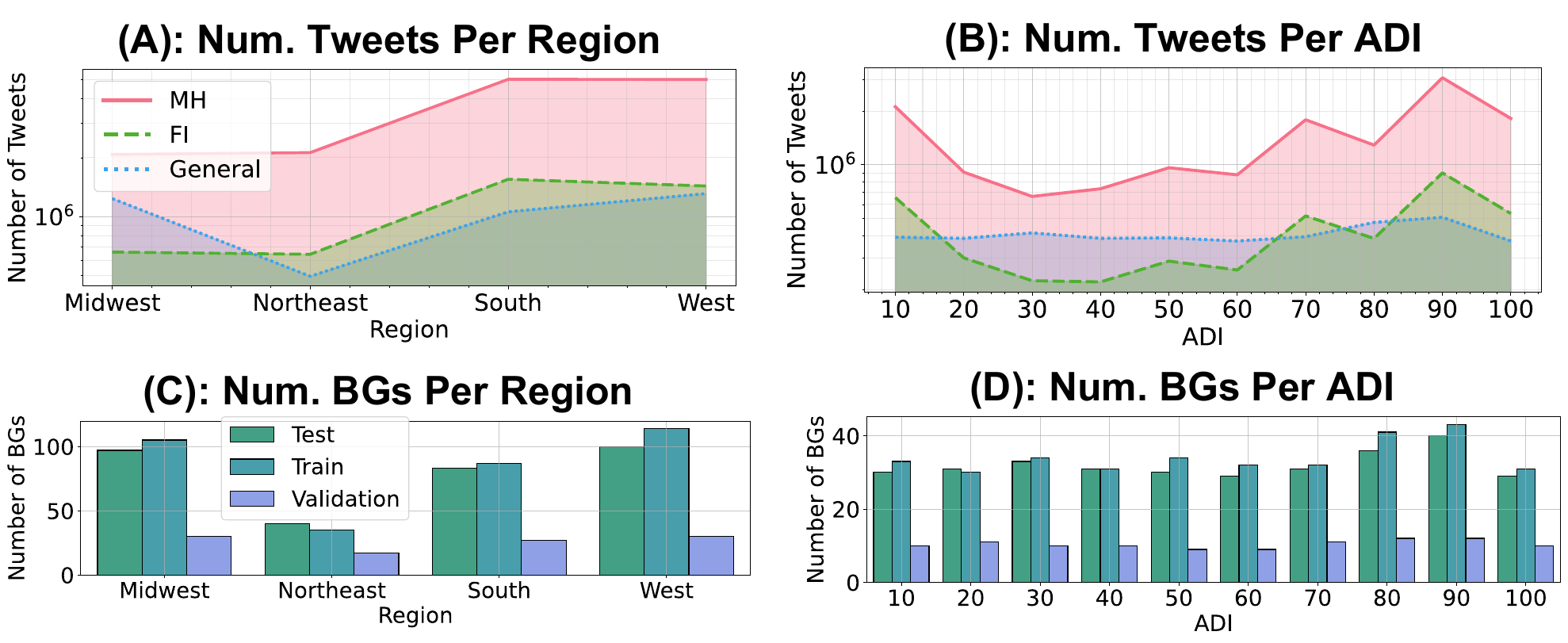}
    \caption{\textbf{Distributional Properties of LocalTweets.}
    (A): Region vs. Number of Tweets: MH tweets are the most numerous, while FI and general tweets have comparable volumes. Tweet volume is slightly skewed toward the South and West regions. (B) ADI vs. Number of Tweets: MH and FI tweets are slightly skewed toward ADIs $\geq 70$ and $\leq 20$. (C) Region vs. Number of BGs: The distribution of data splits over regions is approximately the same. The number of BGs from the Northeast region is less than other regions. Refer to Appendix \ref{appendix:regional_dist} for more discussion. (D): ADI vs. Number of BGs: The number of BGs is fairly balanced over the ADI values and across the data splits.}
    \label{fig:data_properties}
\end{figure*}

Numerous studies evaluated Twitter data for surveillance applications \cite{greco2023transformer, mavragani2020infodemiology, jordan2018using, proserpio2016psychology, klein2022toward, kim2023analyzing, de2016discovering, abdellaoui2017filtering, simonsen2016infectious, shakeri2021digital}. Twitter-based surveillance studies can be divided into three categories: (1) \textbf{Correlation Studies:} studies that investigate the agreement between Twitter data and reported cases of health conditions \cite{paul2011you, broniatowski2013national, coppersmith2015adhd, velardi2014twitter, paul2015worldwide, schwartz2013characterizing, culotta2014estimating, jashinsky2014tracking}; (2) \textbf{Tweeet/User-level Studies:} studies that develop tweet-level or user-level categorization systems to identify tweets or users relevant to a particular health condition \cite{braithwaite2016validating, de2017gender, coppersmith2014quantifying, coppersmith2015adhd, barbieri2020tweeteval, naseem2022benchmarking, yadav2020identifying}; and (3) \textbf{Population-level Studies:} methods that process a set of tweets to make population health inferences. The setup of our study is closest to the third type. 

\textbf{Population-level Studies:}
Recently conducted studies by \citet{barbieri2020tweeteval, naseem2022benchmarking} show that the population-level inference tasks are not present in the current evaluation benchmarks for Twitter data-based Natural Language Processing (NLP) systems. However, there are a few notable studies. 
In studies conducted by \citet{culotta2014estimating, schwartz2013characterizing, giorgi-etal-2018-remarkable}, the authors encode tweets using a count-based system and then use the encodings to make county-level inferences.  
In studies conducted by \citet{nguyen2016building, nguyen2016leveraging, nguyen2017social}, the authors develop machine learning systems that can extract essential indicators of health from tweets and show that extracted indicators are associated with state or census tract-level health outcomes. \citet{athanasiou2023long} conducted clustering analysis and encoded tweets to represent the presence of word clusters. The authors later used encoded tweets along with other data sources for country-level prediction of influenza-like illness outcomes.  
In a recent study conducted by \citet{zhang2022intelligent}, the authors first developed a tweet-level identification system to focus on COVID-related tweets and then used the filtered pool to make country-level COVID outcome prediction. In the above-mentioned studies, either in the data collection or in the encoding process, a focus is put on a specific set of keywords or features. In addition, the authors consider large geographic areas (the smallest area being the census tract in \citet{nguyen2016building}) for making predictions.  

\section{Data}
For the presented analysis, we collected tweets posted from 1,000 census block groups in the United States. We refer to a block group as a neighborhood and use both terms interchangeably. Then, we coupled the Twitter data with mental health outcome estimates reported by the Center for Disease Control (CDC). We refer to the final cleaned version of the collected data as the LocalTweets dataset. In the following subsections, we discuss our data collection process in detail.

\subsection{Sampling of Block Groups}
We started by sampling 1,000 block groups (BGs) from the contiguous United States. Specifically, we stratified the BGs by geographic region (northeast, south, midwest, west) \cite{region2023def, region2023def2}, and Area Deprivation Index (ADI) \cite{adi2023def} \footnote{ADI values are calculated based on the data collected in the American Community Survey (ACS) 5-year estimates at Census Block Group level,   representing the socio-economic profile of a respective block-group. ADI values are between one to a hundred, the highest value being the most undesirable.}. We created 40 strata (four regions and ten ADI bins) and sampled 25 BGs from each stratum.

\subsection{Collection of Twitter data}
\label{section:querying_twitter_data}

Identification and collection of tweets that are relevant to the mental health (MH) status (i.e., outcome of interest) of a population is challenging. Population MH status is an expansive construct influenced by a multitude of demographic, socio-economic, infrastructural, and other factors. Thus, for Twitter data collection, it may not be possible to create a set of keywords that exhaustively cover all possible linguistic expressions of MH-related distress across demographic and cultural features of the population. Hence, we hypothesize that unfiltered tweets may present better datasets for population-level MH surveillance tasks. However, to test our hypothesis we collect keywords-based filtered tweets as well. Overall we collect three subsets of Twitter data, each subset corresponding to a unique category of tweets. The categories are defined by three mutually exclusive sets of keywords used for the collection of tweets. First is the MH category i.e., a subset of tweets that contain keywords directly related to the outcome of interest (MH outcome in our case). For the second category, we list keywords that map to a risk factor of the outcome of interest. In our analysis, we focus on food insecurity, a well-proven contributor to MH \cite{pourmotabbed2020food, elgar2021relative}. Lastly, we collect ``general'' tweets i.e., tweets that contain only one keyword, a space character. We provide our manually curated keyword lists for MH, FI, and general categories in Appendix \ref{appendix:data_collection}. 
Using the Twitter Developer API and twarc \cite{ed_summers_2023_7799050} library, we collect all three subsets of tweets from BGs selected in stratified sampling. For general tweets, we upper bound our collection to 1,000 tweets per BG per year due to the sheer volume of general tweets. We repeat this process for each year from 2015 to 2019. We also collect counts of tweets for each category without any upper bound, using the ``twarc count'' command. Additional details of the twarc library parameters are provided in the Appendix \ref{appendix:data_collection}.

\subsection{Coupling the data with ground truth labels}
\label{section:coupling_data}

After collecting tweets for each sampled BG, we pair the Twitter data with health outcomes published by CDC (also referred to as ground truth) \cite{centers2022places}. For the presented study, we primarily focus on one outcome, the percentage of the adult population with MH not good for more than 14 days of the last 30 days \cite{places2023def}. These prevalence estimates are reported annually at the BG level. Hence, for every unique pair of a year and a BG, we couple the collected set of tweets with the prevalence estimate value available from CDC data. In Figure \ref{fig:data_collection_flowchart}, we provide a schematic of our data collection process.

\subsection{Cleaning and splitting the data for model development}
\label{section:data_splits}

We filter the Twitter-CDC coupled data to remove BGs with no tweets (for any category and any year), and BGs not included in the CDC data. This filtering process removed 235 BGs, resulting in a final dataset of 765 BGs with corresponding Twitter and CDC data for five years. We refer to the cleaned version of the data as LocalTweets. Figure \ref{fig:data_properties} shows the distributional properties of the LocalTweets. We provide a detailed description of data properties in Appendices \ref{appendix:data_properties} and \ref{appendix:longitudinal_properties}. In the cleaned version of LocalTweets, we observe significantly fewer BGs from the Northeast region. Hence, we analyzed the effect of fewer BGs from the Northeast region on the space generalizability of the data. We observe that fewer BGs from the Northeast region do not hamper the space generalizability of the data and provide details in Appendix \ref{appendix:regional_dist}.

We split the data differently for our two experimental settings namely, the forecasting setting and spatial extrapolation setting. For forecasting, we use all 2019 data as the test dataset and consider 2015 to 2018 data for model development. For the spatial extrapolation, we first divide the final set of 765 BGs into three splits (test, train, and validation) such that the distribution over ADI values and geographic regions is approximately held the same across splits, refer to Figure \ref{fig:data_properties}, panels C and D. We remove the BGs in test split, 
to create a proxy set of "unreported" BGs. We use 2015-19 data from the train and validation split for model development and report the performance on the unreported BGs (test split) of the year 2019. 


We denote LocalTweets as $D$, defined as follows:

\begin{equation*}
    D = \{(t^{(k, s)}_{b, y}, c^{(k, s)}_{b, y}, g^{(s)}_{b, y}, r^{(s)}_{b, y})\}
\end{equation*}

where the subscripts b and y represent the BG and year, the superscripts k and s denote the tweet category and split. The variables t, c, and g represent the tweets, tweet counts, and ground truth outcome values, respectively. We create a risk category variable $r$, such that $r^{(s)}_{b, y} = 1$ (high-risk BG) if $g^{(s)}_{b, y} \geq 75^{th}$ percentile, otherwise $r^{(s)}_{b, y} = 0$ (low-risk BG). In other words, $r$ is nothing but a flag indicating whether a BG is high-risk or not. We leverage the variable $r$ to evaluate regression models' performance with discrete metrics (e.g., accuracy and F1-score).
Note that the superscript k is not present for $g$ and $r$ variables because we only consider the MH outcome as the target variable. 

\section{Methodology} 
Our methods of analysis are mainly divided into two parts. First, we conduct a correlation analysis to investigate the agreement between Twitter data-based statistics and the reported CDC outcome values. In the second part, we present a regression analysis. Specifically, we develop a model that processes the Twitter data (set of tweets) to predict the MH outcome value for respective neighborhoods,
refer to Appendix \ref{appendix:localhealth_schematic} for a simple schematic. 
In the following subsections, we discuss each part in detail.

\subsection{Correlation analysis}
We use the Pearson correlation coefficient \cite{pearson2023def, 2020SciPy-NMeth, pearson2023scipy} to measure the correlation between the ground truth values $g_{b,y}$ (the reported MH outcomes) and the Twitter activity i.e., tweet counts  $c_{b,y}^{(MH)}$, $c_{b,y}^{(FI)}$ and $c_{b,y}^{(General)}$. In addition, we also measure the correlation between the ground truth and the ADI values. The correlation is measured separately for each year and over the 765 BGs in the LocalTweets.

\subsection{Regression analysis}
\label{section:regression_analysis}
In the regression analysis, our goal is to develop a parametric function that can predict the continuous and scalar-valued MH outcome for a BG ($g_{b,y}$), based on the set of tweets posted from the same BG. We develop two types of models, one utilizing the tweet count values and the other utilizing the set of tweets. For the count-based model, we utilize the normalized count values (normalized by the count of all tweets posted from the BG) for the MH and FI categories of tweets. We adopt a simple linear regression setting for the count-based model, where normalized count values act as input variables. For the case when we consider both, MH and FI counts, we treat each normalized count value as a separate variable in the linear regression model.

For the text-based model, we follow Algorithm \ref{table:algorithm}, with four main steps, sampling, encoding, aggregation, and prediction. In the sampling step, if the total number of tweets exceeds the 4K mark, we uniformly sample 4K tweets. The sampled tweets are encoded with a language model and then aggregated across sequence length and number of tweets. Finally, we employ a convolutional neural network ($f_{conv}(\cdot)$) followed by a fully connected neural network ($f_{fcn}(\cdot)$) to predict the MH outcome value based on aggregated encodings ($\bar{v}^{(k)}_b$), therefore,

\begin{equation}
    \hat{g}_b = f_{fcn}(f_{conv}(\bar{v}^{(k)}_b; \theta_{conv}); \theta_{fcn})
\end{equation}


\begin{algorithm}[ht!]
    \SetKwInput{Input}{Input}
    \SetKwInOut{Step}{Step}
    \SetKwInOut{Require}{Require}
    \SetKwInOut{Case}{Case}

    \hrulefill
    
    \Input{D, $f_{\text{LM}}(\cdot)$, $f_{\text{conv}}(\cdot)$, $f_{\text{fcn}}(\cdot)$}
    \Require{$B = \{\text{BGs in LocalTweets}\}$, \\ 
    $Y = \{2015, \dots, 2019\}$ \\ 
    }

    \hrulefill
  
    \Step{\textbf{Sampling}}
  
    \For{each $b \in B$, $y \in Y$}{
        Sample $t_{b_s,y}^{(k)} \sim \text{Uniform}(t_{b,y}^{(k)})$
    }
    Such that, $|t_{b_s,y}^{(k)}| = \min(4000, |t_{b,y}^{(k)}|)$ 
    $\forall \ b, y, k$

    \Step{\textbf{Encoding}}
    \For{each $b \in B$, $y \in Y$}{
    $v_{b_s,y}^{(k)} = f_{\text{LM}}(t_{b_s,y}^{(k)};\theta_{\text{LM}})$\;
    }
    
    Such that, $v_{b_s,y}^{(i)} = [v_{b_s, y, 1}^{(k)}, v_{b_s, y, 2}^{(k)}, \ldots, v_{b_s, y, n}^{(k)}]$, where $v_{b_s, y, j}^{(k)}$ is the representation vector of the $j^{th}$ tweet in the $t_{b_s,y}^{(k)}$ set and $n = |t_{b_s,y}^{(k)}|$

    \Step{\textbf{Aggregation}}

    \For{each $b \in B$, $y \in Y$}{
    $\bar{v}_{b_s, y}^{(k)} = \frac{1}{|t_{b_s,y}^{(k)}|} \cdot \sum_{j=1}^{|t_{b_s,y}^{(k)}|} v_{b_s, y, j}^{(k)}$
    }
    
    \vspace{10pt} 
    \Step{\textbf{Prediction}}

    
    $\hat{g}_{b, y} = f_{\text{fcn}}(f_{\text{conv}}(\bar{v}_{b_s, y}^{(k)};\theta_{\text{conv}});\theta_{\text{fcn}})$

    

    

    

    \hrulefill

  \vspace{10pt} 
  \caption{\textbf{LocalHealth Approach.} In this table, we present the LocalHealth algorithm to predict mental health outcome values based on a set of tweets. There are four main steps namely,  sampling, encoding, aggregation, and prediction of outcome value. Superscript $k$ denotes the tweet category.}
  \label{table:algorithm}
\end{algorithm}

Where, $\theta_{conv}$ and $\theta_{fcn}$ represent parameters of the $f_{conv}(\cdot)$ and $f_{fcn}(\cdot)$, respectively. We predict the reported MH outcome based on various categories of tweets by simply changing the $\bar{v}^{(k)}_b$. When we consider both, MH and FI tweets to make predictions ($k=\{MH\ \text{and}\ FI\}$), we add the vectors for MH and FI to compute the final aggregated vector i.e., $\bar{v}^{(k)}_b = \bar{v}^{(MH)}_b + \bar{v}^{(FI)}_b$. We refer to our approach presented in Algorithm \ref{table:algorithm} as LocalHealth.

\section{Experimental Setup}
\label{section:experimental_setup}

\subsection{Sets of experiments}
We divide our experiments into four sets. 

\textbf{Set-1: Effect of input information type.}
\label{section:exp_set_1}
In this set of experiments, we compare the effects of different information priors (ADI values, tweet counts, tweet texts, and tweet categories) on forecasting MH outcomes. Hence, we use data from 2015 to 2018 for developing the model 2019 data for testing (forecasting data splits discussed in Section \ref{section:data_splits}). We also augment the count-based and text-based models with ADI information to measure the impact of combined information. To augment ADI, we concatenate the normalized ADI value ($ADI/100$) with the scalar output of LocalHealth ($f_{fcn}(f_{conv}(\bar{v}))$) and pass the vector to a linear layer to predict the target outcome value.

\textbf{Set-2: Effect of text encoder.}
\label{section:exp_set_2}
In this experiment, we replace the language model ($f_{LM}(\cdot)$) in Algorithm \ref{table:algorithm} with various pre-trained language models, including Twitter-RoBERTa \cite{barbieri2020tweeteval}, PHS-BERT\cite{naseem2022benchmarking}, and GPT-3.5 \cite{openai-gpt3.5-doc}, and assess the changes on forecasting performance. We work with ``general'' category of tweets and consider 2015 to 2018 data for developing the model and, 2019 data for testing. We also measure the zero-shot performance of GPT-3.5, refer to Appendix \ref{appendix:zero-shot} for further details.

\textbf{Set-3: Effect of data availability.}
\label{section:exp_set_3}
In the first two experiments, we utilize data from 2015 to 2018 for training and validating the model. Here in the third set, we gradually reduce the data availability from the prior four years (2015-18) to the prior year (2018) and examine the changes in the forecasting performance subject to data availability. Specifically, we create four train and validation sets based on the data from 2015 to 2018, 2016 to 2018, 2017 to 2018, and only 2018, respectively. We keep the test set (2019 data) constant for all data availability scenarios and compare two language models in this setting, RoBERTa-base, and GPT3.5.

\textbf{Set-4: Spatial extrapolation capabilities.}
\label{section:exp_set_4}
In the fourth set of experiments, we evaluate the LocalHealth approach based on the capabilities to extrapolate CDC outcomes to the unreported BGs (refer to Section \ref{section:data_splits}). 
We consider five data availability scenarios, 2015 to 2019, 2016 to 2019, and likewise till 2019-only. Based on data availability we vary the training and validation data while keeping the test data fixed to 2019 data for proxy unreported BGs. 


\begin{table}

\centering
\begin{tabularx}{\linewidth}{ccccc}
\toprule
\textbf{Year} & \textbf{MH} & \textbf{FI} & \textbf{General} & \textbf{ADI} \\
\midrule

2015 & 0.1640 & 0.1460 & 0.1299 & 0.6767 \\
2016 & 0.1366 & 0.1332 & 0.1215 & 0.7074 \\
2017 & 0.1123 & 0.1132 & 0.0969 & 0.7257 \\
2018 & 0.0928 & 0.0937 & 0.0863 & 0.7162 \\
2019 & 0.0922 & 0.0954 & 0.0832 & 0.7318 \\

\bottomrule
\end{tabularx}

\caption{\textbf{Correlation Results.} In this table, the columns MH, FI, General, and ADI represent the Pearson Correlation Coefficient between the CDC-reported MH outcome i.e., $g_{b_s, y}$ and count of mental health, food insecurity, general tweets, and the ADI values, respectively. All correlation coefficients are statistically significant with $p < 0.05$.}
\label{table:results_correlation}

\end{table}

\subsection{Language models and regression head}
In Set-1 experiments, we use the RoBERTa (base configuration) model \cite{liu2019roberta} to encode the tweets. In Set-2, we provide results for multiple language models e.g., Twitter-RoBERTa \cite{barbieri2020tweeteval}, PHS-BERT \cite{naseem2022benchmarking}, GPT-3.5 \cite{openai-gpt3.5-doc}, etc. We do not update the parameters of the pre-trained language model. Only the parameters corresponding to the regression head of our framework i.e., $\theta_{conv}$ and $\theta_{fcn}$, are updated. For all our experiments, we fixed the structure of the convolutional head to have a single channel, a kernel size of 16, and a stride of four. 

\subsection{Baseline models}
We provide four baseline models. First is the majority baseline i.e., predicting all BGs in the test set as non-high-risk BGs ($r_{(b, 2019)} = 0,\ \forall \ b$). Second, we report the performance of the linear regression model making predictions based only on normalized ADI values. For the third and fourth baselines, we make use of Logistic Regression (LoR) and Support Vector Machine (SVM) models, respectively, along with aggregated general tweet encodings ($\bar{v}$) (from RoBERTa-base model) to directly predict the risk category ($r$) of the BGs. 

\subsection{Hyperparameters and evaluation}

We train all models for 1,600 epochs with a batch size of 512. We use a linear learning rate schedule with a 20\% warmup and peak learning rate of $1\times10^{-3}$. We minimize mean squared error (MSE) using AdamW \cite{loshchilov2017decoupled, adamw2023pytorch} with a weight decay of 0.1. In order to leverage standard classification metrics for model evaluation, we employ a thresholding technique to convert the continuous-valued model outputs ($\hat{g}_{b,y}$) into binary risk category predictions. This allows us to directly compare predicted risk categories with the ground truth labels (variable $r$, Section \ref{section:data_splits}) using established metrics like accuracy and F1-score (macro-averaged). Model selection is conducted based on the macro-F1 score achieved on the validation set. Following training, we evaluate the best model on the test split and report averaged macro-F1 and accuracy across 10 random seeds. For both, SVM and LoR, we change the loss function to binary cross-entropy and we use a classification threshold of 0.15 to identify high-risk BGs.

\section{Results}

We started the assessment of the surveillance utility of Twitter data with correlation tests, between the reported MH outcomes and tweet counts.  We find that tweet counts moderately correlate with the MH outcome, refer to Table \ref{table:results_correlation}. The correlation strength for general tweet counts is consistently lower than the MH or FI tweet counts. In other words, a higher volume of general tweets moderately correlates with worse MH outcomes but, the higher count of MH or FI tweets correlates with worse outcomes marginally better.
For validation, we also conducted a correlation test between MH outcome and ADI value and confirmed much higher correlation strengths compared to all categories of tweets.
To further scrutinize the utility of Twitter data, in the latter part of our analysis, we conducted four sets of experiments to evaluate the LocalHealth approach. We will discuss the results of each set of experiments one by one.


\begin{table}

\centering
\begin{tabularx}{\linewidth}{Xccc}
\toprule
\textbf{Input information} & \textbf{F1-score} & \textbf{Acc. (\%)} \\
\midrule
\textbf{Majority baseline}  & 0.4336 & 76.56 \\
\textbf{Text-based (LoR)} & 0.5224 & 52.75 \\
\textbf{Text-based (SVM)} & 0.5510 & \textbf{76.73} \\
\textbf{ADI-only (LR)} & 0.6406 & 72.81 \\

\midrule
\multicolumn{3}{p{7cm}}{\textbf{Count-based (LR)}} \\ 
MH only & 0.5052 & 61.95 \\
FI only & 0.4465 & 67.88 \\
MH and FI & 0.5133 & 63.84 \\
General only & -- & -- \\

\midrule
\multicolumn{3}{p{7cm}}{\textbf{Text-based (LocalHealth)}} \\ 
MH only & 0.5668 & 60.05   \\
FI only & 0.5602 & 64.43   \\
MH and FI & 0.5853 & 64.16    \\
General only & 0.5984 & 66.76   \\

\midrule
\multicolumn{3}{p{7cm}}{\textbf{Count-based (LR) with ADI}} \\ 
MH-only & 0.5647 & 69.36 \\
FI only & 0.5545 & 71.57 \\
MH and FI & 0.6138 & 69.31 \\
General only & -- & -- \\

\midrule
\multicolumn{3}{p{7cm}}{\raggedright{\textbf{Text-based (LocalHealth) with ADI}}} \\ 
MH only & 0.7089 & 74.52 \\
FI only & 0.7117 & 75.36 \\
MH and FI & 0.7085 & 74.33 \\
General only & \textbf{0.7236} & 76.48 \\

\bottomrule
\end{tabularx}

\caption{\textbf{Effect of Input Information Type.} Here, we present the F1-score and accuracy (Acc.) for identifying the risk category of BGs. Within text-based models, general tweets present better performance than other tweet categories. LoR and LR stand for logistic and linear regression, and SVM for support vector machine.}
\label{table:results_exp_set_1}

\end{table}

\textbf{Set-1: Effect of input information type.}
In the first set of experiments, we compared various information priors available in LocalTweets: tweet count, tweet texts, tweet categories, and ADI values. 
The count-based regression models failed to exceed any of the baselines except the majority baseline, refer to Table \ref{table:results_exp_set_1}. The text-based LocalHealth models performed better than the count-based models but fell short of the ADI baseline.  The F1-score improved significantly, on average by 18\% for count-based models and by 24\% for text-based models, after augmenting with ADI values. Notably, the text-based model augmented with ADI outperformed the individual counterparts, with an F1-score of 0.7236 and an accuracy of 76.48\%, refer to Table \ref{table:results_exp_set_1}. This result highlighted the complementary nature of the information contained in tweets compared to the ADI values.

For text-based models, a comparison within the tweet categories revealed interesting insights. The model with general tweets performed better than other categories of tweets. This finding supports our hypothesis (refer to Section \ref{section:querying_twitter_data}) and highlights the better generalization capabilities of the general tweets for population-level MH outcome prediction, compared to the keyword-derived tweets. 
Hence, our finding motivates the usage of general tweets for the prediction of population-level MH outcomes. We provide additional comparison of the statistical properties of tweet categories in Appendix \ref{appendix:tweet_categories}.


\begin{table}[t!]

\centering
\begin{tabularx}{\linewidth}{@{}Xccc@{}}
\toprule
\textbf{Language Model} & \textbf{Train} & \textbf{F1-} & \textbf{Acc.} \\
                         & \textbf{Par.} & \textbf{score} & \textbf{(\%)} \\
\midrule
Majority baseline        & --   & 0.4336 & 76.56 \\
GPT3.5 (0-shot)          & 0   & 0.4675 & 76.21 \\
ADI only                 & 2   & 0.6406 & 72.81 \\

\midrule
RoBERTa-base             & 210  & 0.7236 & 76.48 \\
RoBERTa-large            & 274  & 0.7228 & 76.04 \\
Twitter-RoBERTa-base     & 210  & 0.7245 & 76.44 \\
PHS-BERT                 & 274  & 0.7301 & 76.97 \\
GPT3.5                   & 402  & \textbf{0.7429} & \textbf{79.78} \\
\bottomrule
\end{tabularx}

\caption{\textbf{Effect of Text Encoder.} Here, we present the F1-score and accuracy (Acc.) for identifying the risk category of BGs. RoBERTa presents competitive results compared to the best-performing GPT3.5. Train Par.: Trainable parameters in LocalHealth setting.
}
\label{table:set_2}

\end{table}

\textbf{Set-2: Effect of text encoder.}
By focusing our attention on the text-based model (general tweets) augmented with ADI information, we measure the effect of changes in the language model ($f_{LM}(\cdot)$) used for encoding tweets.

We experiment with five language models, RoBERTa-base, RoBERTa-large, Twitter-RoBERTa-base, PHS-BERT, and GPT3.5, and present our results in Table \ref{table:set_2}. The effect of the size of the language model was mixed. We observed a minor reduction of 0.0008 in the F1-score for the RoBERTa-large compared to the RoBERTa-base. For the domain-adapted models, we observed an increment of 0.0056 in F1-score for the PHS-BERT (250 mil. parameters) compared to Twitter-RoBERTs (120 mil. parameters). The effect of domain adaptation was consistent for various sizes of the language models. Twitter-RoBERTa improved F1-score and accuracy by 1.2\% and 1.0\% compared to RoBERTa-base. The same improvements were 1\% and 0.6\% for PHS-BERT compared to RoBERTa-large. 
Interestingly, we observed a striking 59\% improvement in the F1-score for GPT-3.5 compared to GPT3.5 in a zero-shot setting. This highlights the difficulty of the MH outcome prediction task, especially for the zero-shot setting. Of all language models evaluated, we observed the best F1-score and accuracy of 0.7429 and 79.78\% for the GPT-3.5 model when used in the LocalHealth approach. 

\begin{figure}[t!]
    \centering
    \includegraphics[width=0.45\textwidth]{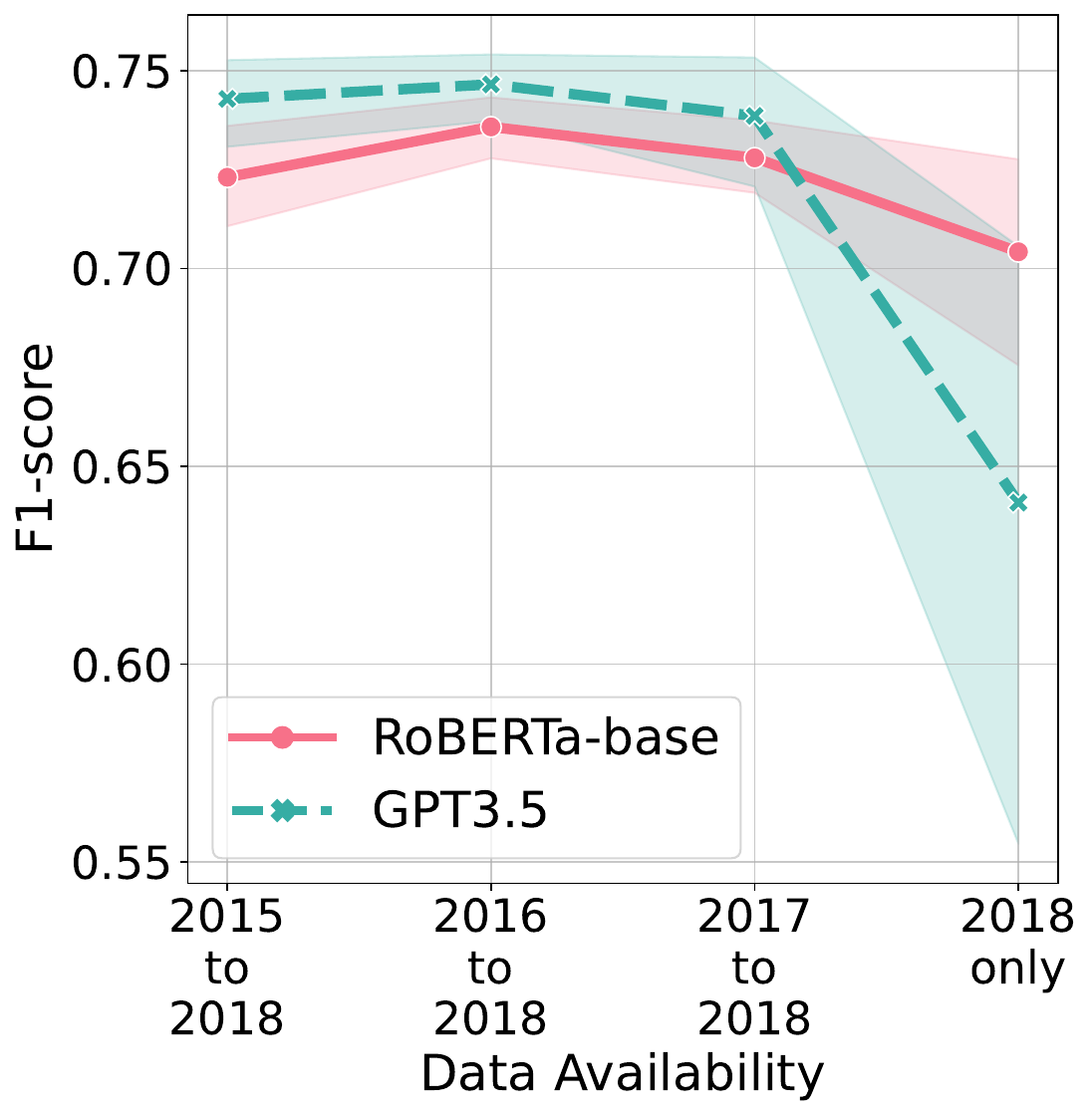}
    \caption{\textbf{Effect of Data Availability on Prediction of Future Outcomes.} In the figure, we present the effect of data availability (x-axis) on the prediction of future i.e., 2019 (all 765 BGs in LocalTweets), MH outcomes. We evaluate models on the correct identification of the BG risk category and plot the F1-score on the y-axis. The lines and shaded regions represent the average value and range of F1-scores, calculated over 10 seeds.}
    \label{fig:exp_3_data_reduction}
\end{figure}

\textbf{Set-3: Effect of data availability.}
In this set of experiments, we investigate the impact of varying data availability on the performance of the models. Performance trends for both models, GPT-3.5 and RoBERTa-base, reveal valuable insights.

Contrary to our expectations, we observed a slight declination in the F1-score when we augmented the 2015 data with the data from 2016 to 2018 for developing the model, refer to Figure  \ref{fig:exp_3_data_reduction}. The declination is 0.0036 for GPT3.5 and 0.0127 for RoBERTa-base. This unexpected dip may be attributable to the possibility that Twitter posts in 2015 may not accurately represent the population MH status in 2019. Interestingly, we found that RoBERTa (with an F1-score of 0.7042) outperforms GPT-3.5 (F1-score: 0.6406) when only 2018 data is available to develop the model. In other cases, when more data is available, GPT-3.5 presents as a better choice of tweet encoder. To this end, we compared the statistical properties of RoBERTa-base and GPT-3.5. We observe a higher variance in the encodings taken from the RoBERTa model and speculate that the more spread out data potentially helps LocalHealth model to detect underlying patterns for MH prediction (refer to Appendix \ref{appendix:longitudinal_properties}). 

\begin{figure}[t!]
    \centering
    \includegraphics[width=0.45\textwidth]{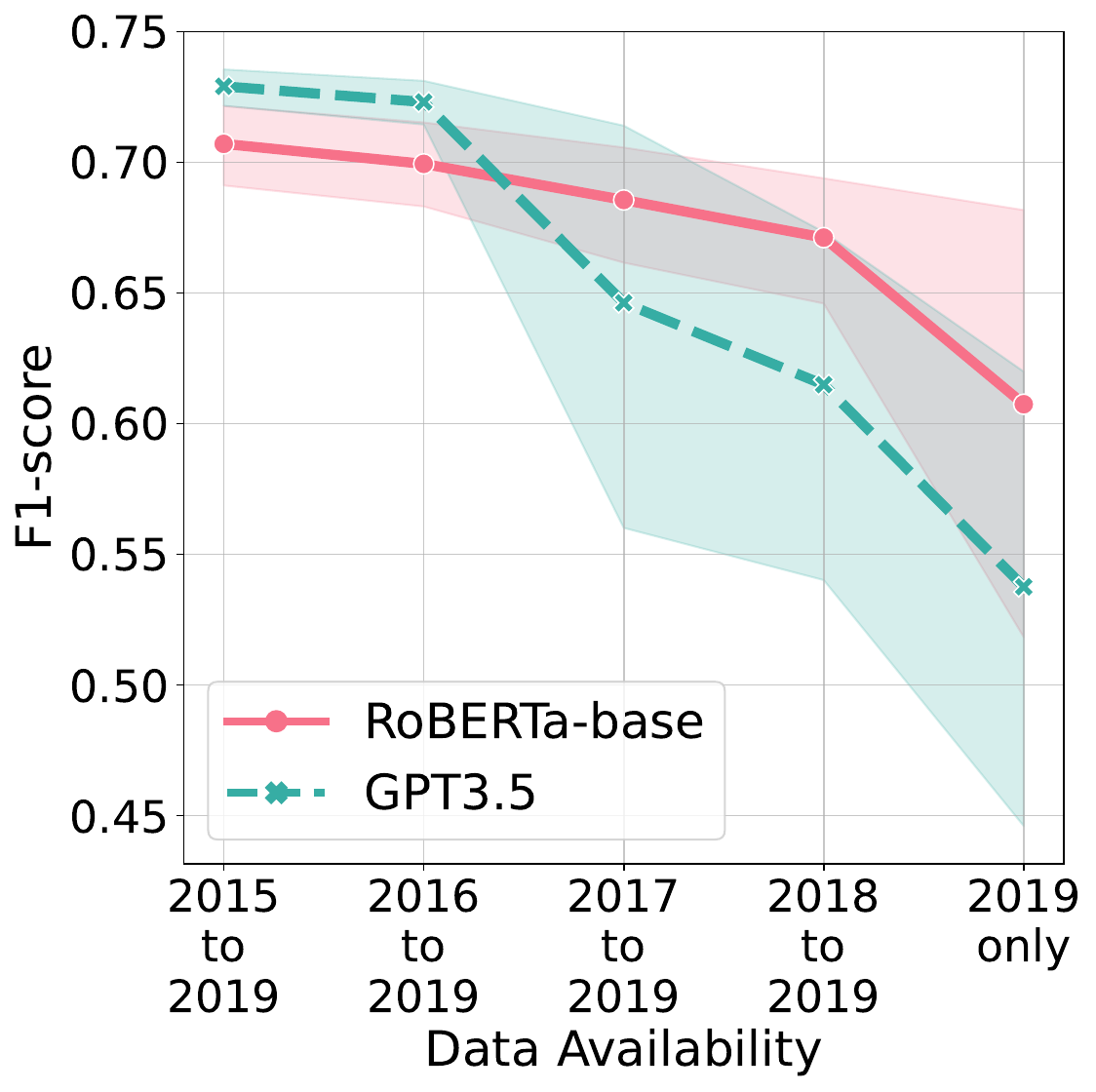}
    \caption{\textbf{Effect of Data Availability on Prediction of Outcomes for Unreported Neighborhoods.} In the figure, we present the effect of data availability (x-axis) on the prediction of MH outcomes for a set of proxy unreported BGs (test split in 2019, 320 BGs). We evaluate models on the correct identification of the BG risk category and plot the F1-score on the y-axis. The lines and shaded regions represent the average value and range of F1-scores, calculated over 10 seeds.}
    \label{fig:extrapolation_for_unreported}
\end{figure}

\textbf{Set-4: Spatial extrapolation capabilities.}
In this set of experiments, distinct from the forecasting task in previous iterations, our focus shifts to predicting MH outcomes for unreported neighborhoods (BGs). Opposed to the findings of Set-3 experiments (forecasting task), we observe that more data is always beneficial for both, RoBERTa-base and GPT3.5, refer to Figure \ref{fig:extrapolation_for_unreported}. Because the proxy set of unreported BGs is never seen by the model, training on more data likely helps the model to find generalizing patterns. Similar to the findings of Set-3, we observe that for limited data availability, RoBERTa stands out as a superior choice of text encoder. In addition, we find RoBERTa to be more robust to changes in the data availability compared to GPT-3.5. The F1-score values for various data availabilities span over a narrow range of 0.0997 ($[0.6074, 0.7071]$) for RoBERTa, while the same range is almost double; 0.1915 ($[0.5376, 0.7291]$); for GPT-3.5 (Figure \ref{fig:extrapolation_for_unreported}).

\section{Conclusion and Future Work}

In this study, we introduce LocalTweets, a novel dataset for mental health (MH) surveillance at the neighborhood level, based on locally posted tweets. We present a simple and efficient approach, LocalHealth, to predict health outcomes based on tweets. Our findings suggest that general category tweets generalize better than the tweets filtered with MH-related keywords. Our results also emphasize RoBERTa-base's effectiveness in data-limited settings.

Our work thus lays the groundwork for a more nuanced and responsive approach to population MH surveillance, fostering advancements in natural language processing methodologies. Alongside surveillance, our work can guide public health resource allocation decisions. For example, presented data and methods can be utilized directly to identify neighborhoods that can benefit from the establishment of community health programs.

Extending our analysis, in the future we hope to investigate resource allocation decisions for specific MH and other health conditions. Furthermore, we also plan to broaden our dataset to include a balanced representation of features that impact the care continuum. We believe improvements in this direction can help us understand the care needs of various communities in a better way.

\section{Limitations}
Our study has several limitations that should be taken into account when interpreting the results. 
First, for the stratified sampling of BGs, we do not consider features such as the availability of healthcare facilities in the neighborhood, insurance-holding population, urban-rural status, educational level, etc. As a result, our data may not capture a balanced view of the population along these features that potentially impact health outcomes. 
Second, the tweets collected under the ``general'' category are not randomly sampled due to the chronological ordering of the Twitter data. This may skew the distribution of the data over time of the year and may limit the applicability of our work for seasonal health conditions. 
Third, our framework can not make inferences for the population unable to access the internet or Twitter. However, based on the estimates of  the size of the population not using the internet \cite{meeker2018internet}, we speculate that this limitation only minimally affects our contributions.
Lastly, the cost of our presented framework may increase based on Twitter’s data pricing policy. However, our findings can help users focus on general tweet data and reduce the volume, time, and cost of data collection. 

\section{Ethics Statement}

While this study demonstrates the potential utility of Twitter data for supplementary mental health surveillance, we acknowledge important ethical considerations. In this section, we describe the procedure we adopted to ensure rightful data access, privacy preservation, and gated sharing of the data. 

\textbf{Data Access:}
To access Twitter data we followed the Twitter Developer Account application procedure\footnote{Updated procedure and terms and conditions can be found \hyperlink{https://developer.twitter.com/en/developer-terms}{HERE}.}. Our application for accessing Twitter data was reviewed, scrutinized, and approved by Twitter, based on an academic research proposal focused on leveraging Twitter data for public health applications.

\textbf{Data Privacy Preservation:}
The use of social media data raises privacy concerns. We took rigorous steps throughout our analysis to protect the privacy of the data. Firstly, we selected tweets that are publicly available and did not collect data from any profiles or tweets that are marked private.
For model development, we focused on tweet texts only and did not make use of any additional features of the tweet or the user such as, location or demographic features. We used twarc library to fetch tweets posted from a specific location (BGs) but we did not access or utilize the location feature of the tweets for model development. Furthermore, in the presented study we ensured privacy preservation through encoding and aggregation of the tweets. In the first step textual information gets encoded into high-dimensional vectors and thousands of such vectors are aggregated together for each block-group. With these two steps, we ensure that human-readable text cannot be excavated from the aggregated representation. 

\textbf{Data Bias:}
The final version of LocalTweets likely is biased along demographic characteristics. However, to maintain the privacy of users and ethical usage of the collected data we did not explore bias along demographic features. We assume that the bias exists and we improve the performance of LocalHealth within the constraints of the demographic bias. Nonetheless, we maintain a fairly balanced distribution across ADI values. We clearly show the existing regional bias and conduct detailed analysis to show that it does not affect results majorly. Furthermore, we show that the bias derived from the commonly used keyword-based data collection methods does not generalize well. Hence, we make the best effort to address bias-related issues while maintaining privacy of the data.

\textbf{Usage of LocalHealth:}
We study the usage of LocalTweets and the application of LocalHealth solely as a supplementary system for traditional health surveillance systems. While supplementary surveillance has benefits, it cannot capture lived experiences. Thus our findings should be considered preliminary and complemented by qualitative, participatory research methods. In addition, mental health is a sensitive topic, and care must be taken not to further stigmatize mental illness. While our work aims for the betterment of mental health public policies, we acknowledge that the findings of our study could be used to develop algorithms that can target distressed areas or populations at risk with discriminatory or harmful content. Hence, we will provide gated access to LocalHealth. We will release the model and the data based on
individual requests that adhere to, (1) focus usage of the data for research on public health research questions (2) follow Twitter's data privacy policy.

\textbf{Reproducability:}
Lastly, on the technical front, we made our best efforts to reduce the technical barrier to research by considering economic language models, and training lean ($\leq402$ parameters) systems. Our goal was to encourage community participation for the benefit of the community. However, we recognize that the barrier is also contingent upon Twitter's privacy policies.

\nocite{*}
\section{Bibliographical References}\label{sec:reference}

\bibliographystyle{lrec_natbib}
\bibliography{lrec-coling2024-example}

\bibliographystylelanguageresource{lrec_natbib}
\bibliographylanguageresource{languageresource}

\appendix

\section{Collection of Tweets}
\label{appendix:data_collection}

Twitter data collected in this study was retrieved using the Twitter Developer API. Specifically, we leveraged the twarc library \cite{ed_summers_2023_7799050} for querying data from Twitter. Multiple query features were used to retrieve the required data. Out of all the features, the list of keywords and location information were vital for our analysis. The lists of keywords were curated separately for mental health and food insecurity-related tweets, refer to Table \ref{table:appendix_keywords}. For the general category of tweets, we use only one space character to retrieve tweets. The location information feature was used to retrieve Tweets from a specific geographic area. We considered the census block groups as the geographical unit for the collection of Tweets. The centroid of a block group along with a radius value was used to define the block group area in the twarc data retrieval query. We varied the radius value based on the population density of the county that a block group belongs to. We primarily employ the variability in the radius value to focus on an area that is appropriately scaled according to the demographics. For example, in the locations with low population density a fixed radius may focus on a very small area and hence, we would not be able to collect any tweets. Likewise for a densely populated area if the fixed radius is set to too high a value, then we may collect an eccentrically high volume of tweets. Here, we assume that the population density ($\rho$) is uniform for any county, therefore,

\begin{equation*}
    \rho_{county} = \rho_{BG \in county} = \frac{population_{BG}}{\pi \cdot r^2}
\end{equation*}

We find the variable radius value as follows,

\begin{equation*}
    r = \sqrt{\frac{population_{BG}}{\pi \cdot \rho_{county}}}
\end{equation*}

We collect the BG population values from \citet{manson2020ipums} and county density values from \citet{county2023density}. Lastly, to avoid very large or very small values of the radius we keep an upper and a lower bound of 10 and 2 miles, respectively.


\begin{table*}[t] 
    \centering
    \begin{tabular}{cp{0.75\textwidth}}
        \toprule
        \multicolumn{1}{c}{\textbf{Category}} & \multicolumn{1}{c}{\textbf{Keywords}} \\ 
        \midrule
        Mental health &
        'bored', 'disgusting', 'sick of', 'tired of it', 'dont want to', 'so fucking miserable', 'tired of being', 'depressed', 'alone', 'isolate', 'given up', 'no friend', 'cant deal', 'want to talk', 'in my room', 'awake', 'sleepless', 'nightmares', 'insomnia', 'cant sleep', 'wish sleep', 'up all night', 'body is begging', 'exhausted', 'tired', 'my energy', 'dont have energy', 'tired to look', 'feel myself falling', 'binge', 'fasting', 'eating disorder', 'eat again', 'always eating', 'forced to eat', 'am eating ?', 'failure', 'ugly', 'worthless', 'hate myself', 'fat piece', 'self hatred', 'piece of shit', 'feel like trash', 'thoughts', 'confused', 'overthinking', 'am losing', 'losing mind', 'my mind off', 'quiet', 'attention', 'nervous', 'social anxiety', 'dead quiet', 'dont wanna move', 'cut', 'hang', 'blade', 'die', 'suicidal', 'rip skin', 'suicide attempt', 'car hit', 'kill myself', 'of the road' \\
        \midrule
        Food insecurity &
        "food stamps", "SNAP", "food charities", "food pantry", "food voucher", "deficiency", "hunger", "hungry", "food insecurity", "poor diet", "junk food", "food desert", "poor nutrition", "starvation", "without food", "no food", "no groceries", "lack of food", "not enough food" \\
        \midrule
        General & 
        " " \\
        \bottomrule
    \end{tabular}
    \caption{\textbf{Keywords for Collecting Tweets}. We use manually curated keywords for specific categories `mental health', `food insecurity', and `general' category of tweets. For the `general' category, we use a space character as the only keyword.}
    \label{table:appendix_keywords}
\end{table*}

\begin{figure}[t!]
    \centering
    \includegraphics[width=0.4\textwidth]{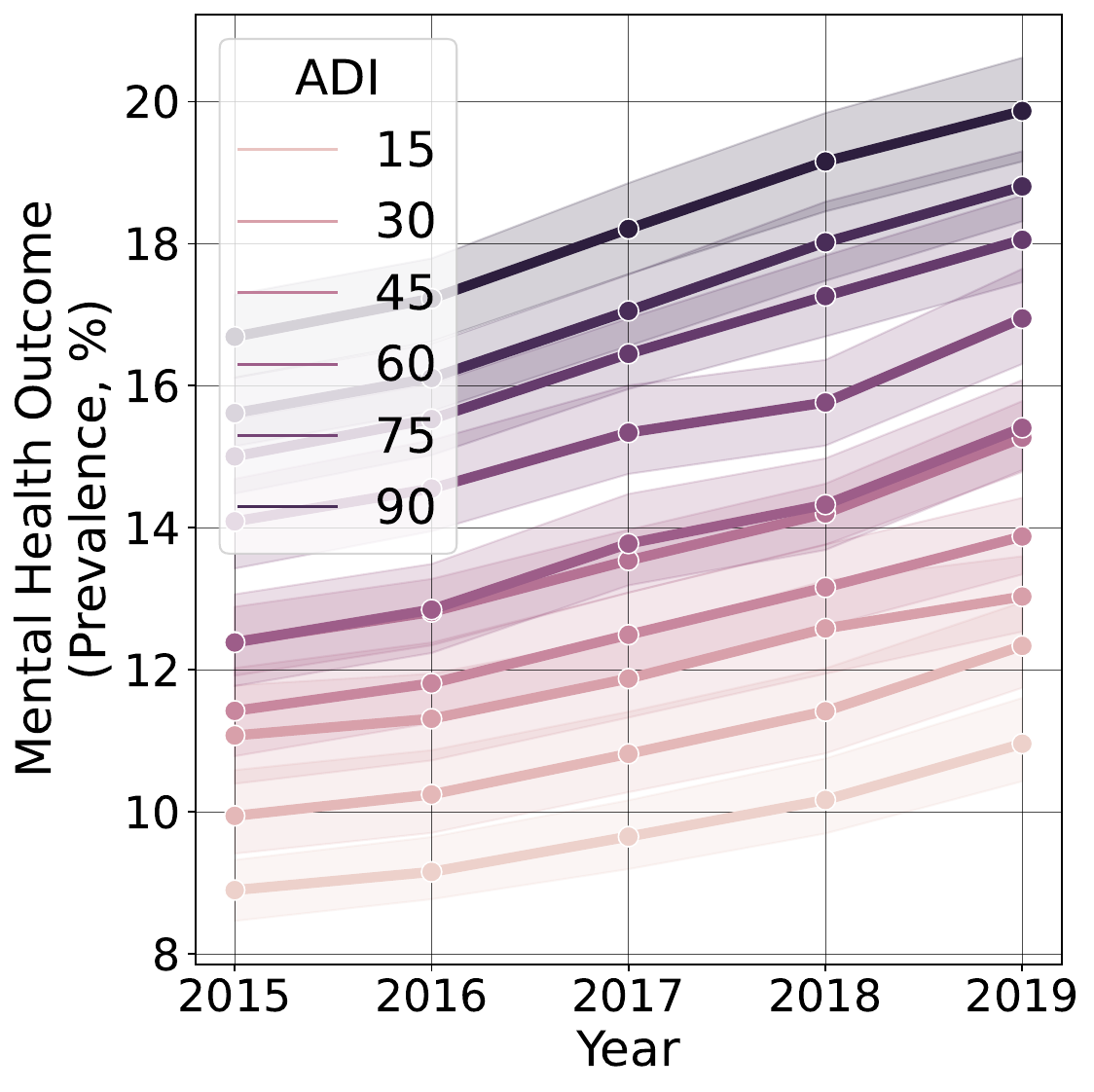}
    \caption{\textbf{Longitudinal Trend of Target Variable.} In this figure, we present the average values of MH outcomes for the years 2015 to 2019. The average value is calculated separately for each ADI value considered in the analysis. The increasing trend in the MH outcome values is observed across all BGs irrespective of their socio-economic status.}
    \label{fig:longitudinal_target}
\end{figure}

\begin{figure}[t!]
    \centering
    \includegraphics[width=0.4\textwidth]{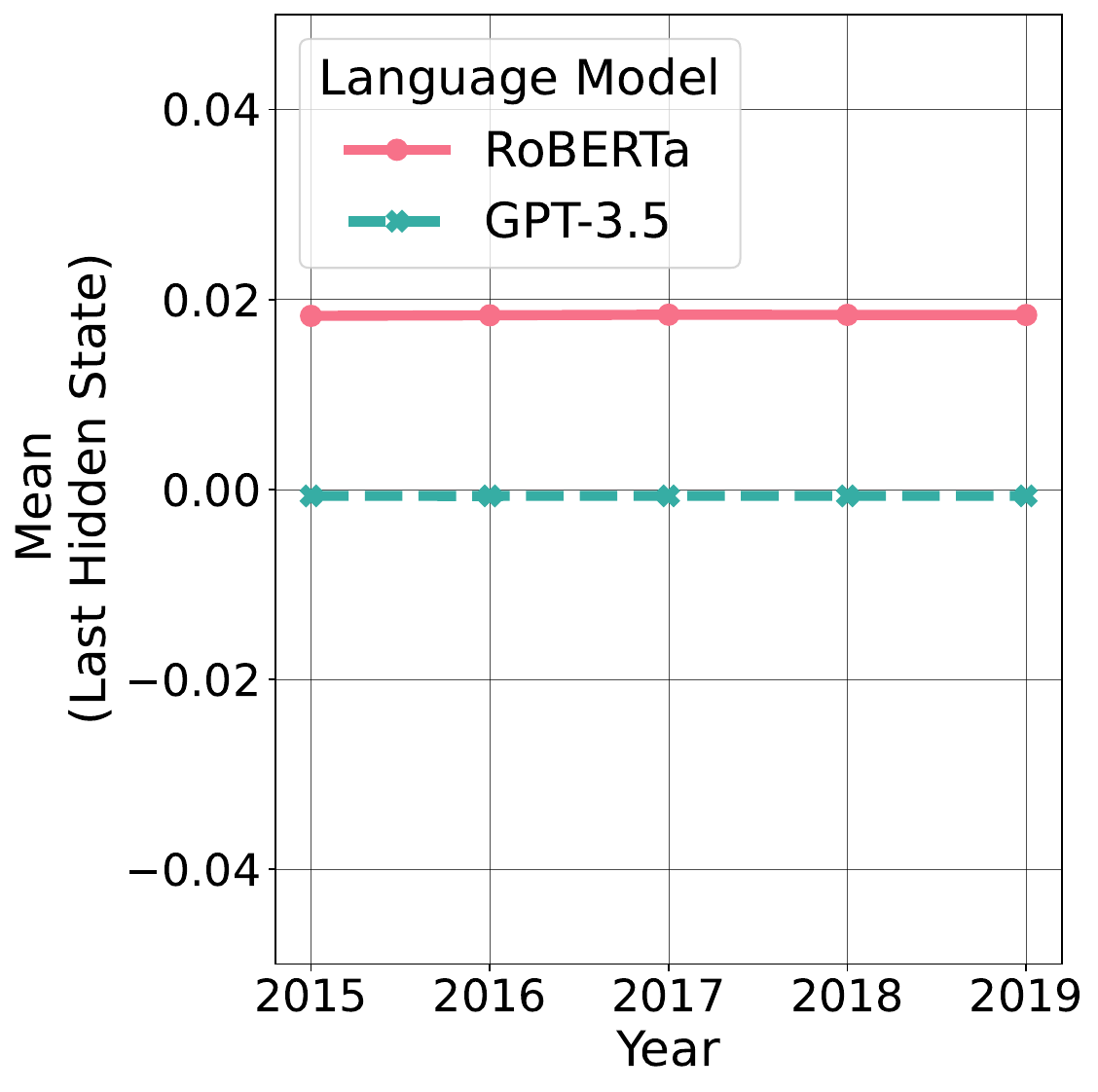}
    \caption{\textbf{Average of Activations.} In this figure we present the average value of the last hidden state for RoBERTa and GPT-3.5 model. The average values are calculated separately for each year, over all dimensions of the hidden vector and 765 block-groups. We observe fairly stable average values, primarily due to the normalization operations included in the language models.}
    \label{fig:longitudinal_lm_average}
\end{figure}

\begin{figure}[t!]
    \centering
    \includegraphics[width=0.4\textwidth]{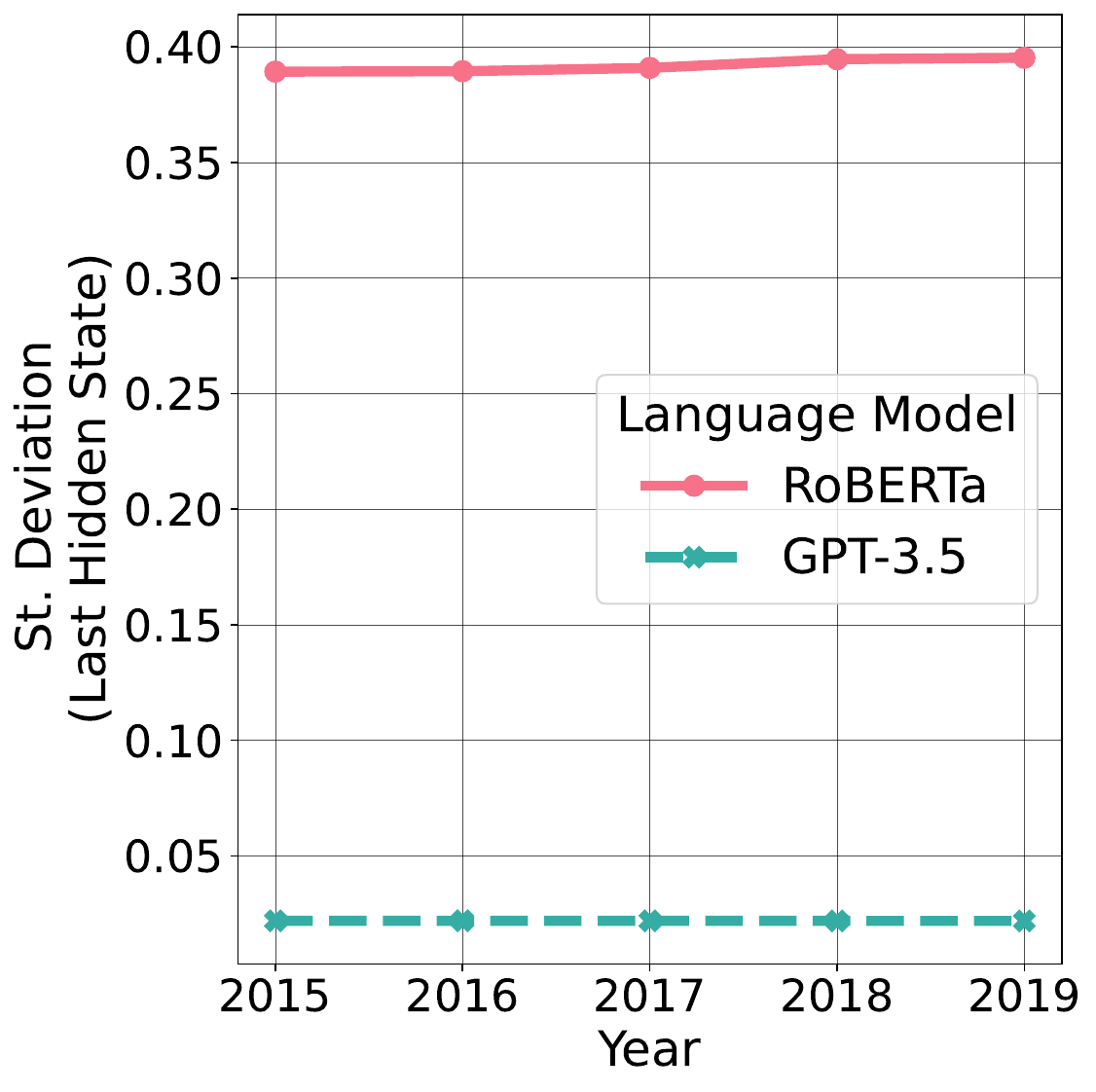}
    \caption{\textbf{Standard Deviation of Activations.} In this figure we present the standard deviation (STD) of the last hidden state for RoBERTa and GPT-3.5 model. The STD values are calculated separately for each year, the overall dimensions of the hidden vector, and 765 block-groups. We observe a significant difference between the STD values for RoBERTa and GPT-3.5. In addition, we also note a slight increase in STD for RoBERTa in the year 2018 and 2019.}
    \label{fig:longitudinal_lm_std}
\end{figure}

\section{Data Properties}
\label{appendix:data_properties}
We calculate two main statistics for setting a few parameters in our analysis, the number of words per tweet and the number of tweets per block group (BG). For calculating the number of words per tweet, we simply count whitespace-separated words and use the distribution of this statistic to guide our decision of setting sequence length for encoding tweets. First, we calculate percentiles of the number of words per tweet separately for each year and tweet category, refer to Table \ref{table:twitter_data_properties} for the values. We consider the maximum value of $75^{th}$ percentile i.e., 29 (for MH tweets in 2019), and estimate the number of ByteBPE \cite{liu2019roberta} tokens per tweets, as $29 \times 1.32 = 38.28$ based on the findings presented by \citet{deshpande2023honey}. Finally, we set the sequence length parameter to the next power of 2 i.e. 64, for encoding tweets (encoding step in Algorithm \ref{table:algorithm}). Similarly, we calculate percentile values for the number of tweets per BG and set the tweet sample size upper bound for each BG equal to 4,000, making sure we cover all $75^{th}$ percentile values. Lastly, we present the distribution of the MH outcome values reported by CDC in the Table \ref{table:twitter_data_properties}. We defined risk categories for BGs based on the distribution of the reported MH outcome values.  Using the $75^{th}$ percentile value for each year we set the status of the BGs as high-risk if the reported outcome value is more than the $75^{th}$ percentile value. 


\begin{table*}[t]
\centering
\begin{tabular}{@{}ccccccc@{}}
\toprule
{\textbf{Tweet Category}} & {\textbf{Year}} & \multicolumn{5}{c}{\textbf{Percentiles}} \\
 \cmidrule(r){3-7}
& & 0 & 25 & 50 & 75 & 100 \\
\midrule

\multicolumn{7}{c}{\centering{\textbf{Number of words per tweet}}} \\

\midrule
\midrule

FI & 2015 & 1 & 6 & 11 & 16 & 93 \\
FI & 2016 & 1 & 7 & 11 & 17 & 89 \\
FI & 2017 & 1 & 8 & 13 & 18 & 176 \\
FI & 2018 & 1 & 8 & 14 & 23 & 179 \\
FI & 2019 & 1 & 9 & 15 & 25 & 196 \\
\midrule
MH & 2015 & 1 & 6 & 11 & 18 & 122 \\
MH & 2016 & 1 & 7 & 12 & 18 & 104 \\
MH & 2017 & 1 & 8 & 14 & 20 & 163 \\
MH & 2018 & 1 & 9 & 16 & 28 & 202 \\
MH & 2019 & 1 & 9 & 17 & 29 & 191 \\
\midrule
General & 2015 & 1 & 7 & 11 & 16 & 106 \\
General & 2016 & 1 & 8 & 12 & 16 & 91 \\
General & 2017 & 1 & 8 & 12 & 17 & 206 \\
General & 2018 & 1 & 8 & 13 & 22 & 172 \\
General & 2019 & 1 & 8 & 13 & 23 & 216 \\

\midrule

\multicolumn{7}{c}{\centering{\textbf{Number of tweets per BG per year}}} \\

\midrule
\midrule

FI & 2015 & 20 & 207 & 496 & 1,229 & 25,841 \\
FI & 2016 & 1 & 64 & 274 & 1,135 & 41,595 \\
FI & 2017 & 1 & 46 & 197 & 771 & 29,628 \\
FI & 2018 & 1 & 35 & 160 & 621 & 26,053 \\
FI & 2019 & 1 & 35 & 143 & 539 & 25,403 \\
\midrule
MH & 2015 & 28 & 584 & 1,462 & 3,549 & 88,717 \\
MH & 2016 & 1 & 78 & 591 & 2,878 & 130,157 \\
MH & 2017 & 2 & 58 & 436 & 2,172 & 109,464 \\
MH & 2018 & 2 & 58 & 440 & 2,229 & 121,132 \\
MH & 2019 & 1 & 52 & 370 & 1,854 & 115,160 \\
\midrule
General & 2015 & 1000 & 1,079 & 1,092 & 1,097 & 1099 \\
General & 2016 & 443 & 1,079 & 1,092 & 1,097 & 1099 \\
General & 2017 & 405 & 1,082 & 1,092 & 1,097 & 1099 \\
General & 2018 & 285 & 1,072 & 1,089 & 1,096 & 1099 \\
General & 2019 & 196 & 1,061 & 1,088 & 1,095 & 1099 \\

\midrule

\multicolumn{7}{c}{\centering{\textbf{MH outcome reported by CDC}}} \\

\midrule
\midrule

-- & 2015 & 0.0580 & 0.1010 & 0.1260 & 0.1540 & 0.2300 \\
-- & 2016 & 0.0590 & 0.1040 & 0.1310 & 0.1570 & 0.2320 \\
-- & 2017 & 0.0590 & 0.1120 & 0.1380 & 0.1660 & 0.2610 \\
-- & 2018 & 0.0690 & 0.1160 & 0.1450 & 0.1750 & 0.2820 \\
-- & 2019 & 0.0780 & 0.1240 & 0.1540 & 0.1820 & 0.2910 \\

\bottomrule
\end{tabular}
\caption{\textbf{Distributional Properties of LocalTweets.} In this table, we present the distribution of tweet length, tweet volume, and MH outcome for the data included in LocalTweets. For the properties related to the Twitter data, we present the statistics separately for each category of tweets.}
\label{table:twitter_data_properties}
\end{table*}

\begin{figure}[t!]
    \centering
    \includegraphics[width=0.4\textwidth]{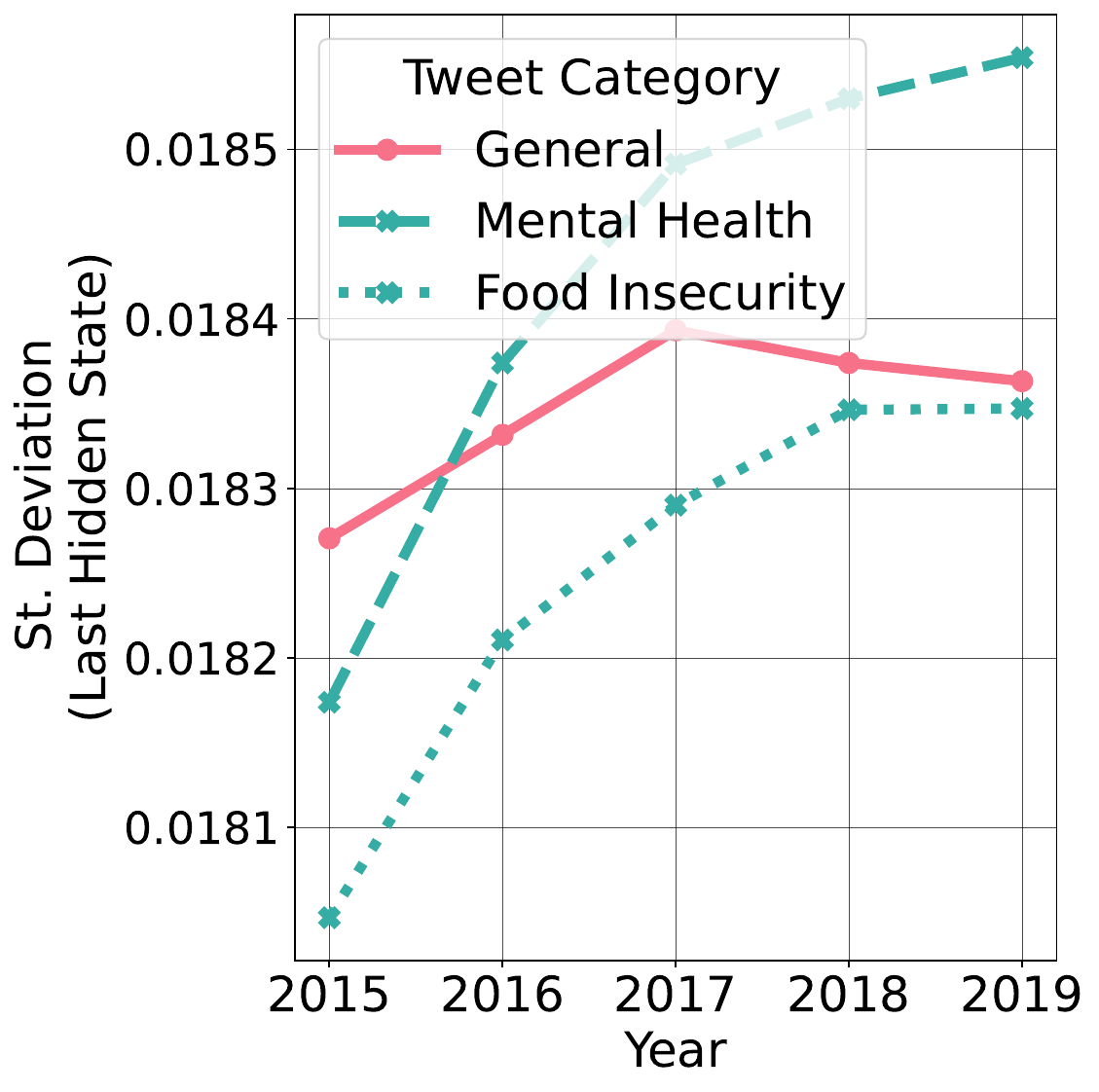}
    \caption{\textbf{Average Across Tweet Categories.} In this figure we present the average of the  RoBERTa last hidden state for all tweet categories. The average values are calculated separately for each year, the overall dimensions of the hidden vector, and 765 block-groups. We observe negligible differences between tweet categories.}
    \label{fig:tweet_categories_mean}
\end{figure}

\section{Longitudinal Properties of Input and Target Variables}
\label{appendix:longitudinal_properties}

In this section, we present a few longitudinal properties of the input and target variables. To reiterate, the input variable is the encoding of a set of tweets from a specific language model (refer to Section \ref{section:regression_analysis}). The target variable is the MH outcomes collected from the CDC database. 

In Figure \ref{fig:longitudinal_target}, we present the variation of the MH outcomes value in time. We calculate average values of MH outcomes across all BGs (765), separately for each year. We observe that MH outcomes have a consistent increasing trend over the years. Interestingly, this trend holds irrespective of the socio-economic status (ADI) of BGs. Hence, to effectively solve the problem of predicting MH outcomes, the model needs to recognize patterns in the tweets that eventually lead to an increasing pattern in the MH outcomes. 

The input variables i.e., the encoding from language model and hence, high dimensional. Hence, to briefly understand the longitudinal patterns in the input variables, we calculate mean and standard deviation values, over all encoding dimensions and BGs, but separately for each year, refer to Figures \ref{fig:longitudinal_lm_average} and \ref{fig:longitudinal_lm_std}. For both language models, the mean values are stable for all years. The mean values for GPT-3.5 are much closer to zero as compared to the RoBERTa model. The reason for such robust mean values is the normalization operations conducted in the language models. While mean values for both models are close to each other, there are noticeable differences in the standard deviation values. The standard deviation of the RoBERTa model is approximately 16 times higher than GPT-3.5. In addition, there also exists a slight increase in the standard deviation value in the RoBERTa model's activations for the year 2018 and 2019. Such increment in the standard deviation values is not observed in the GPT-3.5 model. We believe these statistical properties are the primary reason behind the performance differences between RoBERTa and GPT-3.5 models presented in Figures \ref{fig:exp_3_data_reduction} and \ref{fig:extrapolation_for_unreported}. The high variance of the RoBERTa encodings possibly helps the model to identify underlying hidden patterns effectively. 

\begin{figure}[t!]
    \centering
    \includegraphics[width=0.4\textwidth]{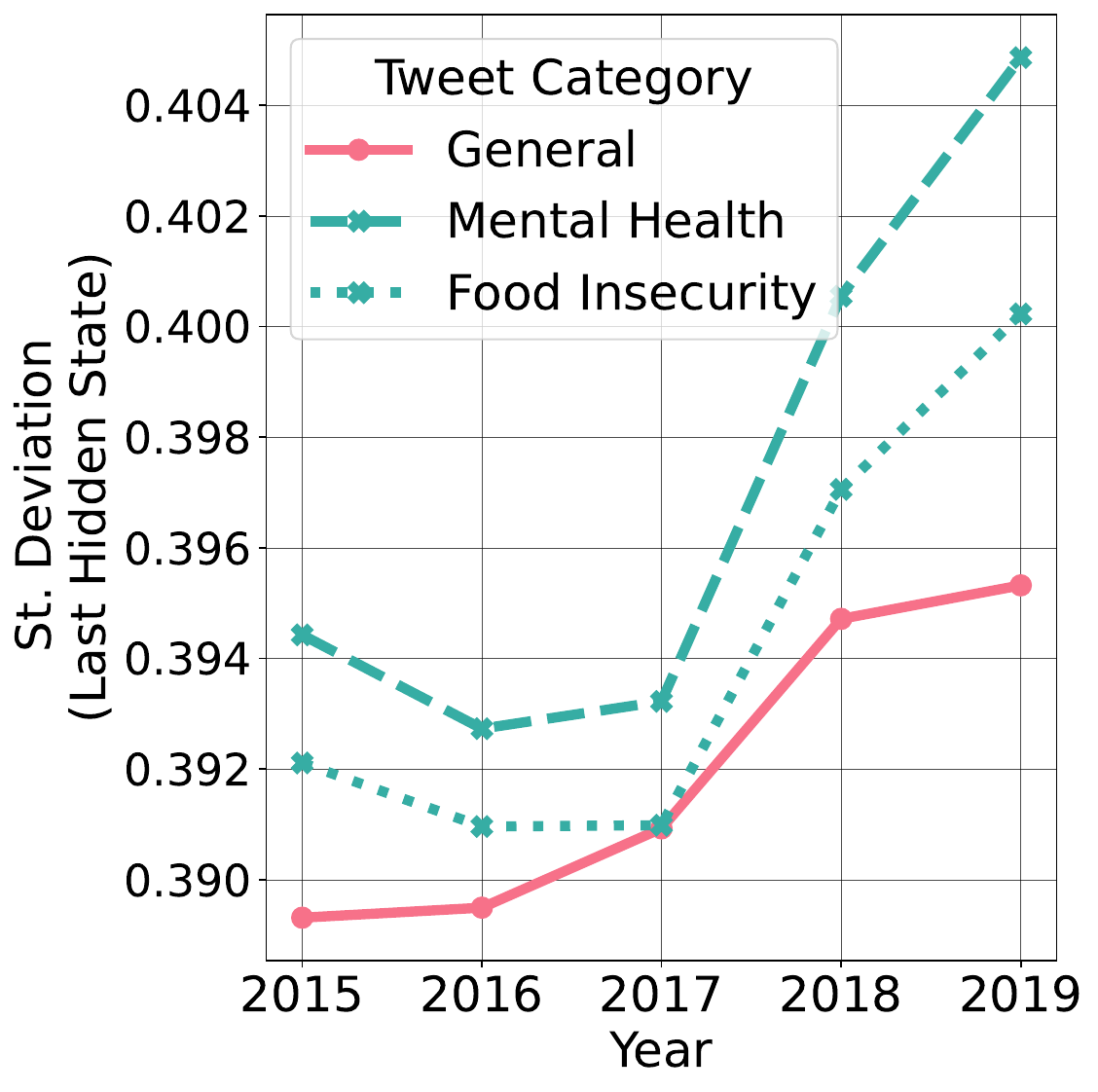}
    \caption{\textbf{Standard Deviation Across Tweet Categories.} In this figure we present the standard deviation (STD) of the  RoBERTa last hidden state for all tweet categories. The STD values are calculated separately for each year, the overall dimensions of the hidden vector, and 765 block-groups. We observe a minor difference between tweet categories. Notably, the variance in the general category tweets is consistently observed as the lowest among all categories, for all years.}
    \label{fig:tweet_categories_std}
\end{figure}

\section{Properties of Tweet Categories}
\label{appendix:tweet_categories}
Our experiment results presented in Table \ref{table:results_exp_set_1} highlight the advantage of using general category tweets over the other mental health (MH) associated categories namely, food insecurity (FI) and MH. Here, we present the mean and standard deviation of RoBERTa encodings for all three categories. In the LocalTweets dataset, we notice only minor differences between the mean and standard deviation for various tweet categories, refer to Figure \ref{fig:tweet_categories_mean}, \ref{fig:tweet_categories_std}. Nonetheless, we observe the standard deviation value for general category tweets to be the lowest, consistently over all years. We speculate that the lower variance in the general category tweets might help focus the model on a specific semantic latent space that is potentially beneficial for the prediction of MH outcomes. 

\section{Effect of Skewed Regional Distribution}
\label{appendix:regional_dist}

In the data cleaning process, we removed the BGs with no tweets or are not reported in the CDC dataset \cite{centers2022places}. In this data cleaning process, the number of BGs included under the Northeast region was reduced from 250 to 94. As a result, the number of BGs from the northeast region is considerably lower compared to other geographical regions in our analysis. Hence, to measure the impact of the skewed distribution we conducted an experiment. We split the cleaned data such that we do not use BGs from the Northeast region to train or validate the model. However, we tested our model on the 2019 data for the northeast region BGs. This setup is similar to our spatial extrapolation setup, refer to Section \ref{section:querying_twitter_data}. We create training data from approximately 75\% of the remaining BGs while using the rest for validation. Otherwise, we keep our experimental setup the same as that of our main experiments, discussed in Section \ref{section:experimental_setup}. We find that the tweets posted from regions other than the Northeast region carry significant generalization capability. With a model trained on general tweets from regions other than the Northeast region, the high-risk BGs in the Northeast region can be identified with an F1-score and accuracy of 0.7450 and 75.09\%. Hence, in conclusion, a skewed distribution over geographical regions will not significantly affect the geographic generalizability of LocalTweets.

\section{GPT3.5 zero-shot experiment details}
\label{appendix:zero-shot}
We conducted a zero-shot performance assessment using GPT3.5-16k \cite{openai-gpt3.5-doc}. Our prompt consisted of the tweets posted from a specific BG and the task was to predict the risk category of the BG. We define the risk category as high-risk for the BGs with mental health outcome values over the 75th percentile. Similar to our experiments with the LocalHealth method, we considered 4,000 tweets for predicting the category of BG. Due to a stringent constraint on the input sequence size (16K tokens), we sampled 100 tweets 40 times instead of feeding 4,000 tweets to GPT3.5-16k. In other words, for each BG, we randomly sampled 100 tweets, with replacement, 40 times and utilized zero-shot prompting, separately for each sampled set. We selected a sample size of 100 based on the median length of the tweets (13, refer to Table \ref{table:twitter_data_properties}) such that, 100 tweets and prompts will fit under the limit of 16,000 tokens. Out of the 40 responses if more than 20 responses categorize the BG as high-risk then we consider the GPT3.5-16k prediction to be high-risk for the respective BG. We adopted the following structure for the prompt:

\begin{lstlisting}[language=Python, style=pythonstyle,linewidth=\columnwidth] 
Tweets: {["tweet1", "tweet2", ..., "tweet100"]}
Instruction: Above tweets are posted from a block-group in the United States in the year 2019. ADI values are representative of the socio-economic profile of a respective block-group. ADI values are between one to a hundred, the highest value being the most undesirable. The ADI index for this block group is {"adi"}
Question: Based on the above tweets and ADI value, what would be the prevalence of adults (>= 18 years) with mental health not good for more than 14 days in a period of 30 days? The range of reported values is from 5% to 30%. The 25th, 50th and 75th percentile values are, 11.1%, 13.9%, and 16.9%, respectively. Select your answer from following options.

Options:
A. High-risk (prevalence greater than the 75th percentile)
B. Low-risk (prevalence less than the 75th percentile)

You must output letter A or letter B

Output:

\end{lstlisting}
Lastly, we spent 215.62 USD in total for the zero-shot experiments with GPT3.5. 

\section{Simplified Schematic for the LocalHealth Method}
\label{appendix:localhealth_schematic}

\begin{figure*}[t!]
    \centering
    \includegraphics[width=1.0\textwidth]{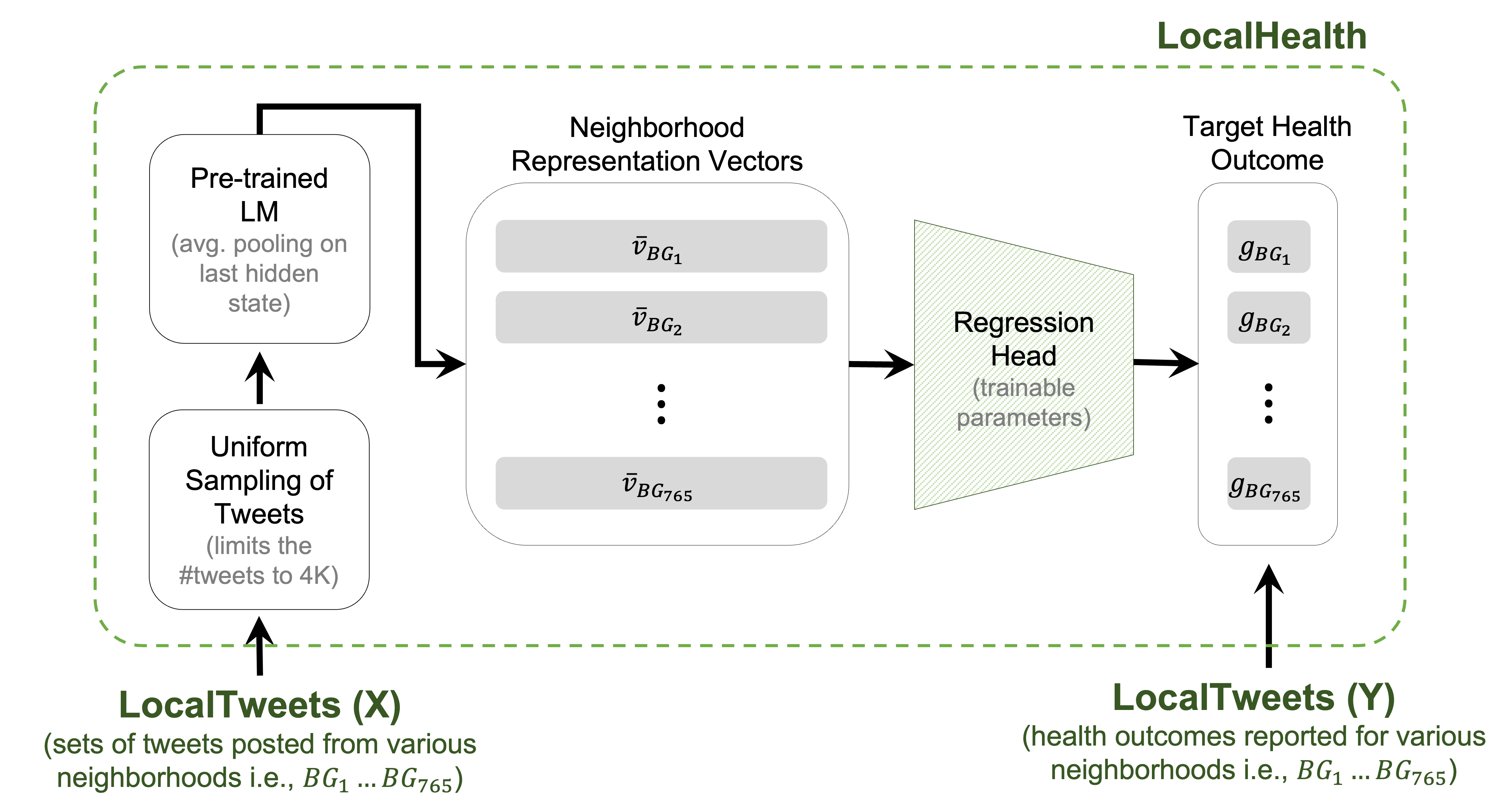}
    \caption{\textbf{LocalHealth Approach.} In this figure, we present a simplified schematic representation of the LocalHealth approach (Algorithm \ref{table:algorithm}).  The input to the LocalHealth method is the set of tweets contained in LocalTweets data. LocalHealth has four main steps: sampling, encoding, aggregation, and prediction. The vector $\bar{v}$ is an aggregated vector representation for individual BGs based on which the MH outcomes are predicted for respective BGs.}
    \label{fig:localhealth_schematic}
\end{figure*}

Our proposed method primarily adopts four steps, sampling, encoding, aggregation, and prediction, as mentioned in the Algorithm \ref{table:algorithm}. In this section we provide a simplified schematic for the Algorithm \ref{table:algorithm}, refer to Figure \ref{fig:localhealth_schematic}.

\end{document}